\documentclass[twocolumn]{aa}
\usepackage{graphicx}
\usepackage{txfonts}
\usepackage{arydshln}
\usepackage{bm}
\usepackage{textcomp}
\usepackage{multirow }
\usepackage{amsmath,graphics}
\usepackage{epstopdf}
\usepackage{soul}

\begin{document}

\title{Frequency rising sub-THz emission from solar flare ribbons}
\titlerunning{Model of the frequency rising sub-THz emission component}
\authorrunning{E. P. Kontar et al}

   \author{E.P. Kontar
          \inst{1}
          \and
          G.G. Motorina \inst{1,2,3}\fnmsep\thanks{E-mail: motorinagalina@gmail.com}
          \and
          N.L.S. Jeffrey\inst{1}
          \and
          Y.T. Tsap\inst{2,4}
          \and
          G.D. Fleishman\inst{5}
           \and
          A.V. Stepanov\inst{2,3}
          }

\institute{School of Physics and Astronomy, University of Glasgow,
           Kelvin Building, Glasgow G12 8QQ, UK
         \and
         Central  Astronomical Observatory at Pulkovo of Russian Academy of Sciences, St. Petersburg, 196140, Russia
          \and
   Ioffe Institute, Polytekhnicheskaya, 26, St. Petersburg, 194021, Russia
           \and
             Crimean Astrophysical Observatory, Nauchny, Crimea
             \and
             New Jersey Institute of Technology, University Heights, Newark, NJ 07102-1982, USA
             }

   \date{Received ; Accepted}

\abstract{Observations of solar flares at sub-THz frequencies
(mm and sub-mm wavelengths) over the last two decades often
show a spectral component rising with frequency.
Unlike a typical gyrosynchrotron spectrum decreasing with frequency,
or a weak thermal component from hot coronal plasma,
the observations can demonstrate a high flux level
(up to $\sim 10^4$~s.f.u. at $0.4$~THz)
and fast variability on sub-second time scales.
Although, many models has been put forward to explain
the puzzling observations, none of them have clear observational support.
Here we propose a scenario to explain
the intriguing sub-THz observations.
We show that the model,
based on free-free emission from the plasma of flare ribbons
at temperatures $10^4-10^6$~K, is consistent with all existing
observations of frequency-rising sub-THz flare emission.
The model provides a temperature diagnostic of the flaring chromosphere
and suggests fast heating and cooling of the dense transition region plasma.}

\maketitle

\section{Introduction}
Solar flares are efficient charged particle accelerators that convert
the energy of the magnetic field into the kinetic energy of electrons and ions.
During their transport, flare-accelerated electrons emit radio waves and hard X-rays (HXRs)
\citep[see][ as recent reviews]{2008LRSP....5....1B,2011SSRv..159..107H}.
The unprecedented X-ray imaging and spectroscopic capabilities of the Ramaty High Energy Solar Spectroscopic Imager \citep[RHESSI; ][]{2002SoPh..210....3L} has improved the diagnostics of non-thermal electrons \citep{2011SSRv..159..301K}, resulting in a clearer understanding
of electron acceleration and transport \citep{2011SSRv..159..107H},
as well as the global energetics of flares \citep{2017ApJ...836...17A,2017PhRvL.118o5101K}.

Since the beginning of this century, it became possible to make sub-mm wavelength observations of solar flares,
at a few frequencies with limited spatial resolution \citep[see][as a review]{2012ASSP...30...61K}. One of the most intriguing aspects of the observations at 0.2-0.4 THz (200-400 GHz) is the presence, in some flares,
of a spectral component that grows with frequency \citep{2001ApJ...548L..95K};
this component is markedly different from the frequency-decreasing gyrosynchrotron spectrum produced by $\sim$~MeV electrons \citep[e.g.][]{1985ARA&A..23..169D}.
While a frequency-growing component due to thermal emission
from a hot coronal plasma is likely to be present in virtually all flares,
the observed flux of $\sim 10^4$~s.f.u. at 400 GHz in some flares
presents a major challenge for standard radio emission models.

Although optically-thick thermal free-free emission
is the first candidate to account for the frequency-rising spectral
component, and often observed in the gradual phases of flares \citep[][as a review]{2013MmSAI..84..405T},
the lack of supporting signatures in other wavelengths triggered concern
about the plausibility of this mechanism \citep[e.g.][]{2007SoPh..245..311S,2010ApJ...709L.127F,2013A&ARv..21...58K}.
The large flux and a noticeable correlation with HXRs led to
the proposal that the emission is likely associated with accelerated non-thermal
electrons \citep{2001ApJ...548L..95K,2009ApJ...697..420K}.
The measurement of radio emission source sizes could provide additional observational constraints regarding the possible emission mechanisms. However, there are currently no reliable radio source size measurements near 400~GHz,
only radio source centroid locations, but they only suggest that the emission is indeed spatially associated with flares \citep{2012ASSP...30...61K}.

There is a long list of emission mechanisms proposed in the literature
\citep[e.g.][]{2007SoPh..245..311S,2010ApJ...709L.127F,2012ASSP...30...61K,2013A&ARv..21...58K,2013AstL...39..650Z}.
Assuming small radio source sizes of $\leq 20''$,
a number of coherent emission mechanisms have been proposed. \cite{2010ApJ...709L.127F} considered radio emission by relativistic electrons
in the chromosphere via Cherenkov emission and the emission from short-wavelength Langmuir turbulence; \citet{2006A&A...457..313S,2013AstL...39..650Z,2014SoPh..289.3017Z} suggested plasma emission in the dense solar atmosphere,
while \citet{2014ApJ...791...31K} proposed that the microbunching instability
plays a key role, and that coherent gyrosynchrotron emission
is produced in the microwave domain,
while the usual gyrosynchrotron emission is in the sub-THz domain.
However, the models proposed have several assumed
conditions and generally suffer from a lack of observational support. Hence, they cannot be verified observationally.
In general, any successful model should explain:
(i) a sub-THz component that increases with frequency \citep{2017SoPh..292...21F},
(ii) a large $\sim 10^4$ s.f.u. flux at 200-400 GHz,
(iii) a close association with non-thermal particles \citep{2009ApJ...697..420K}, and
(iv) the sub-second variations of the radio flux \citep{2009ApJ...697..420K}.

In this paper, we analyse the relationship between the area of flare ribbons
and the flare sub-THz component, and propose a new model based on radio emission from the transition region perturbed by flare-accelerated electron heating.
In such a model, the emission is free-free emission originating
from an optically-thick transition-region plasma with a temperature of $10^4-10^6$~K. Analysing the chromospheric flare ribbons, we show that flares with the largest ribbon area produce the largest sub-THz flux.
The model can predict fluxes based on the transition region temperature
and the flare ribbons can explain flux values for all flares reported in the literature.

\section{Flare observations with sub-THz emission}

\begin{figure*}\centering
\includegraphics[width=0.49\linewidth]{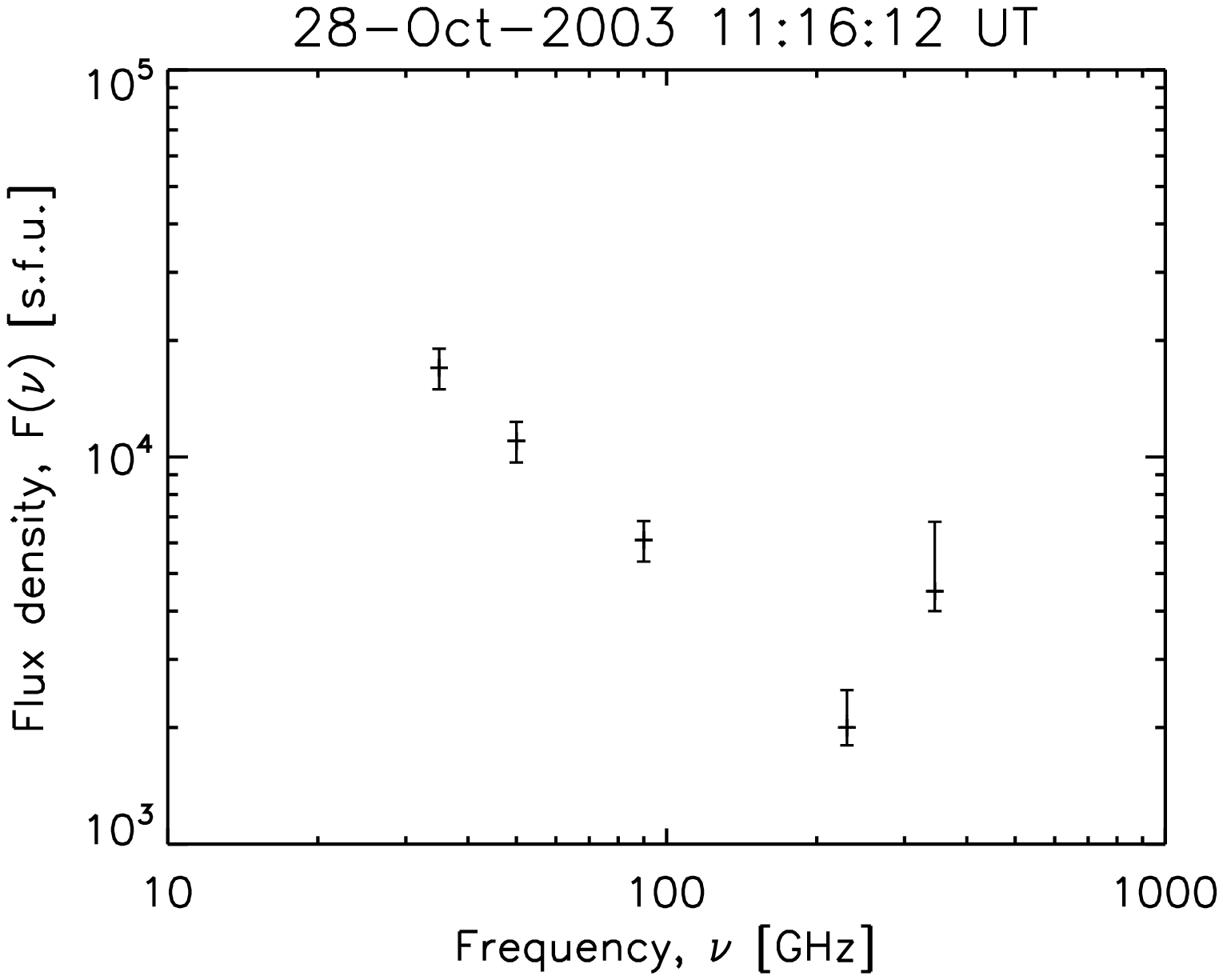}
\includegraphics[width=0.49\linewidth]{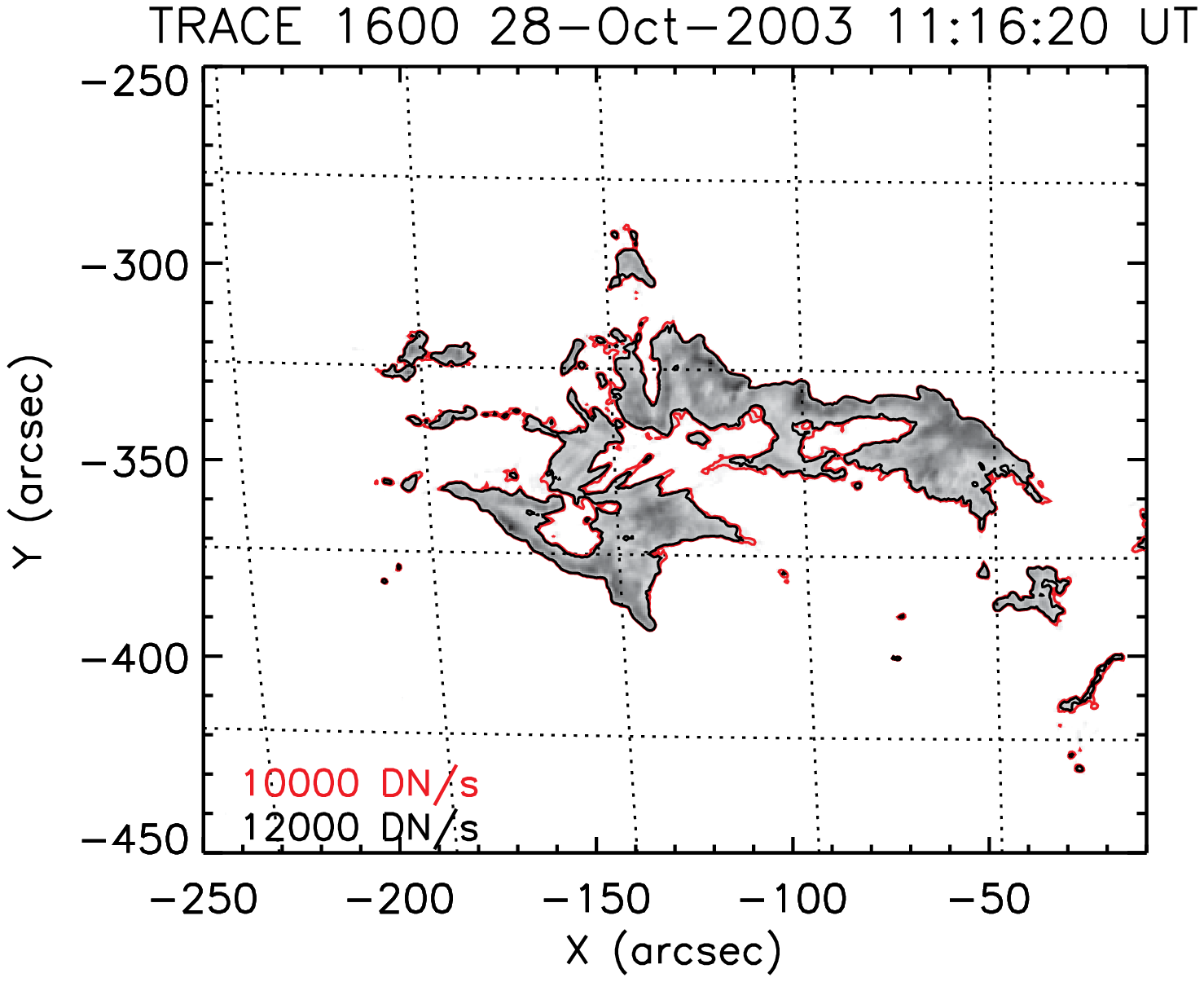}
\caption{\label{fig:ribbons}
For flare SOL2003-10-28T11:10:{\textit{(left)}}
the flux density spectrum $F(\nu)$ showing the rising sub-THz
component  above $200$~GHz, after \citet{2008ApJ...678..509T}
and \textit{(right)} UV solar flare ribbon observed by TRACE in
the 1600~\AA$\space$ passband at 11:16:20 UT; 10000 DN/s and 12000 DN/s levels are shown by red and black contours. }
\end{figure*}

A total of 31 solar flares with radio flux observations at the sub-THz frequency range have been found in the literature: 16 flares have positive spectral slope at the THz frequency range, 12 – negative, 
and 3 – undefined due to absence of observations or large uncertainties. 
Among these 31 flares, we selected 17 for study:
14 flares have a spectrum that grows with frequency
above $100$~GHz (positive spectral slope)
and 3 flares have a frequency-decreasing spectrum (negative spectral slope).
An example of one flare with a frequency-growing sub-THz spectrum is shown in Figure~\ref{fig:ribbons}.
The other 14 flares are not examined due to the absence or unreliability of their data
(i.e. no ultraviolet (UV) observations, $>50$\% flux uncertainties,
low fluxes (<10 s.f.u.), and no radio data at 405 GHz).
All the flares with a frequency-growing sub-THz component are summarised in Table \ref{tab:list1}, while flares with a frequency-decreasing component
are summarised in Table \ref{tab:list2}.
\begin{table*}
\caption{\label{tab:list1} A list of flares with a rising spectrum of sub-THz emission. The solar flare observation date, the GOES flare class, the time of maximum sub-THz
emission, the spectral index, the radio flux density at 405~GHz (where available),
the time the UV ribbon area is measured, and the UV ribbon area, are presented.
Sub-THz radio data are from the following papers:
$^a$\citet{2001ApJ...548L..95K, 2002A&A...381..694T};
$^b$\citet{2004A&A...415.1123L};
$^c$\citet{2011SoPh..273..339T};
$^d$\citet{2004A&A...420..361L, 2008ApJ...678..509T};
$^e$\citet{2007SoPh..245..311S,2009ApJ...697..420K};
$^f$\citet{2004ApJ...603L.121K,2009SoPh..255..131K};
$^g$\citet{2009SoPh..255..131K};
$^h$\citet{2017SoPh..292...21F};
$^i$\citet{2016AdSpR..57.1449T};
$^j$\citet{2018SoPh..293...50T};}
\begin{center}

\begin{tabular}{lcccccc}
\hline \hline
Solar flare (GOES class)& 	Sub-THz (UT)& $\delta$& F. density (s.f.u.)  & UV(UT) &$A_{\rm{UV}}(\rm{arcsec}^2)$  \\
\hline
 SOL2000-03-22T18:48 (X1.1)$^a$
 & 18:50:00&   $1.27_{-0.63}^{+0.63}$ & 500 & 18:50:00  & $412\pm 51$\\
SOL2001-04-12T10:28 (X2.0)$^b$
& 10:17:54 & $1.09_{-1.07}^{+0.96}$ & $810^\dag$ & 10:16:11& $816\pm189$\\
SOL2003-10-27T12:43 (M6.7)$^c$
& 12:32:30  & $1.68_{-0.49}^{+0.48}$ & $86^\dag$ & 12:32:36 & $490\pm143$\\
SOL2003-10-28T11:10 (X17)$^d$
& 11:16:12& $2.0_{-0.53}^{+0.8}$ & $4500^\dag$ &11:16:20  & $3590\pm506$\\
SOL2003-11-02T17:25 (X8.3)$^e$
& 17:19:30 &$3.41_{-1.79}^{+2.92}$& 50000	&17:35:38	& $1268\pm349$ \\
SOL2003-11-04T19:50 (X28)$^f$
& 19:44:00 &	$0.71_{-0.24}^{+0.23}$& 18000 &19:43:11	&$1969\pm390$  \\
SOL2006-12-06T18:47 (X6.5)$^g$
& 18:43:51 &$0.98_{-0.93}^{+0.68}$ & 6800	&18:43:53	& $2785\pm182$\\
SOL2012-10-22T18:51 (M5.0)$^h$
& 18:48:30 &$1.42_{-0.49}^{+0.48}$& 50	&18:49:52	&$148\pm56$ \\
 SOL2012-07-04T09:55 (M5.3)$^i$$^\ddag$
 & 09:55:30 & $1.29^{ +0.6}_{-0.63}$ & 39$^\ddag$ &09:57:04	&$142\pm91$ \\
SOL2012-07-05T11:44 (M6.1)$^j$$^\ddag$
& 11:44:24& $1.33^{+0.6}_{-0.63}$ & 26$^\ddag$   &11:41:52 & $126\pm51$ \\
SOL2013-02-17T15:50 (M1.9)$^h$
& 15:46:25 &$1.42_{-0.28}^{+0.28}$& 200	&15:47:52	&$67\pm27$ \\
SOL2014-10-27T14:47 (X2.0)$^h$
& 14:22:50 &$1.07_{-0.31}^{+0.31}$&60	&14:18:40	&$133\pm53$ \\
SOL2014-11-05T19:44 (M2.9)$^h$
& 19:53:40 &$0.63_{-0.38}^{+0.4}$& 30	&19:53:52	&$15.6\pm15.4$  \\
SOL2014-11-07T17:26 (X1.6)$^h$
& 17:25:30 &$0.52_{-0.29}^{+0.29}$& 70	&17:25:28	&$202\pm99$ \\
\hline
\end{tabular}
\end{center}

$^\dag$~Flux density at 345~GHz, 405~GHz data are not available;

$^\ddag$~Flux density at 140~GHz. The radio spectral index was calculated using 140~GHz and 93~GHz data, as the higher frequencies are not available.

\end{table*}
%\footnotetext{\textbf{WHY IS THIS HERE??? DELETE?}The group of flares, which have a negative/flat/positive spectral slope at sub-THz frequencies during flare evolution.}
\begin{table*}
\caption{\label{tab:list2} A list of flares with a decreasing spectrum of sub-THz emission. The flare parameters are the same as in Table \ref{tab:list1}.
The sub-THz radio data are from $^h$\citet{2017SoPh..292...21F}.}
\centering
\begin{tabular}{lcccccc}
\hline \hline
Solar flare (GOES class) & 	Sub-THz (UT)& $\delta$& F. density (s.f.u.)  & UV(UT) &$A_{\rm{UV}}(\rm{arcsec}^2)$  \\
\hline
%1) SOL2001-08-25T16:45 (X5.3)  &16:31:13   & $-1.06_{-1.1}^{+0.91}$ &5500  &    16:29:02      & $1640\pm563$\\      % inserting body of the table
 %SOL2002-08-30T13:29 (X1.5)  &13:28:10   & $???$ & 20  &    13:07:15    & $211\pm106$\\      % inserting body of the table
%SOL2001-11-28T16:35 (M6.9)  &16:34:01   &$-1.86_{-0.41}^{+0.55}$         & 24 &  16:34:03       & $1068\pm49$\\
SOL2012-01-27T18:37 (X1.7)$^h$
&18:26:00   &$-1.46_{-0.46}^{+0.44}$& 140  &     18:25:53 &  $522\pm163$\\
 SOL2012-03-13T17:41 (M7.9)$^h$
 &17:23:10   &$-1.07_{-0.64}^{+0.54}$ & 100 &       17:19:05& $86\pm44$\\
SOL2014-10-22T14:28 (X1.6)$^h$
&14:06:50  &$-0.28_{-0.36}^{+0.35}$ & 100	&  14:07:28& $299\pm96$\\
\hline
\end{tabular}
\end{table*}
Tables \ref{tab:list1} and \ref{tab:list2} also show the spectral
index $\delta$ between radio fluxes at 212 GHz and 405 GHz,
calculated as
$\delta = \log (F_{\text{405 GHz}}/F_{\text{212 GHz}})/\log(405/212)$,
where $F_{\text{212 GHz}}$ and $F_{\text{405 GHz}}$
are the spectral flux densities at frequencies $\nu=212$ GHz
and $\nu=405$ GHz, respectively, measured at the peak of the sub-THz emission (the exact times used are listed in Tables \ref{tab:list1} and \ref{tab:list2}).
Table \ref{tab:list1} shows that virtually all sub-THz events
have a spectral index of $\delta\le 2$, which is consistent with the assumption
of optically thick free-free emission.
Flare SOL2003-11-02T17:25 has a spectral index of $\delta=3.4^{+2.9}_{-1.8}$,
but the large errors on the spectral index suggest
a spectral index of $\delta=2$ is within the $1\sigma$ uncertainty.
In general, due to the large flux uncertainties \citep[see also][]{2013A&ARv..21...58K},
the spectral index has large $1\sigma$ uncertainties, and hence,
several emission mechanisms including optically thick free-free emission
are consistent with the observed spectral index.

\section{Ultraviolet observations of flare ribbons}

HXR producing non-thermal electrons in flares deposit most of their energy
in the chromosphere \citep[][]{2011SSRv..159..107H}
leading to bright emission from the chromosphere and transition region
and often faint HXR emission in the corona.
Broadband UV images show flare ribbon emission from the transition region and chromosphere \citep[e.g.][]{2001ApJ...560L..87W}. To evaluate the flare ribbon area, we study UV images
at the 1600~\AA  $\space$ passband, from the Transition Region and Coronal Explorer \citep[TRACE; ][]{1999SoPh..187..229H}
for 7 flares before 2010 (the last TRACE science image was in 2010),
and from the Solar Dynamics Observatory Atmospheric Imaging Assembly  \citep[SDO/AIA; ][]{2012SoPh..275...17L} for 10 flares after the year 2010.
Both instruments provide adequate resolution; the TRACE pixel size is $0.5''\times 0.5''$, and the AIA pixel size is $0.6''\times 0.6''$. The time cadence for both instruments varies from 10-12~seconds to several minutes.

Due to the fact that the considered flares are powerful (with GOES classes from M1.9 to X28), the UV images are usually saturated during the flare impulsive phase and the sub-THz emission peak. The saturation issue originates due to
pixels of the charge-couple device (CCD) only being able to accommodate a finite number of counts \citep{1997pgca.book.....M}, and
it can affect the intensity flux estimation. When pixels reach saturation level, spreading to adjacent pixels may begin, causing a secondary saturation (blooming). In turn, the blooming of the images can affect the area estimations.
To recover saturated TRACE and AIA images,
different approaches have been applied. \citet{2001SoPh..198..385L} offer
the Solar SoftWare (SSW) routine  trace\_undiffract\footnote{\url{https://hesperia.gsfc.nasa.gov/ssw/trace/idl/util/trace_undiffract.pro}} to clean TRACE images, while \citet{2015A&C....13..117S} developed the DESAT package to de-saturate SDO/AIA images.
Unfortunately both routines are only valid for a limited number of wavelengths
(for TRACE - only for  171, 195, 284~\AA; for AIA - 94, 131, 171, 193, 211, 304, 335~\AA), and not for 1600~\AA.
In a statistical study of 190 flares observed with TRACE, \citet{2009ApJ...690..347L} define the uncertainty of the outer ribbon edges as a two pixel error.
To infer the ribbon-edge uncertainty,
\citet{2012SoPh..277..165K} defined different ribbon-edge cut-off values from six to ten times the background level.

In our paper, we first define a cut-off range level and subtract
the area of the saturated pixels affected by blooming to estimate
the uncertainty on the flare areas.
To remove saturation effects, the TRACE images have been processed
with the SSW routine trace\_prep.pro,
which corrects for missing pixels, dark pedestal and current,
and replaces saturated pixels with values above 4095 DN.
A similar procedure aia\_prep.pro has been applied to AIA images.
The UV images are analysed at a time when sub-THz emission peaks
or a few minutes before or after the peak if no other images are available or if they are saturated (see Tables \ref{tab:list1}, \ref{tab:list2}). To define the area, we choose two levels which correspond
to the maximum and minimum areas, normalised by the exposure time
intensity, of $I_{min}=2000$~DN/s and $I_{max}=4000$~DN/s respectively.
The areas of the found contours should not be less than 10 arcsec$^2$, to exclude individual pixels and transient non-flare features (as an example see Figure \ref{fig:ribbons}). The UV areas $A_{\rm{UV}}$ that satisfy the above requirements are summarised in Tables \ref{tab:list1} and~\ref{tab:list2}.
Three flares (SOL2000-03-22, SOL2003-10-28, SOL2006-12-06), recorded by TRACE, have minimum and maximum levels of intensity at 10000 and 12000 DN/s respectively,
since the count rate is extremely high.

%The flare SOL2006-12-06T18:47 has minimum and maximum levels of intensity at  %$I_{min}=65000$~DN/s and $I_{max}=75000$~DN/s respectively,

\section{UV flare ribbon areas and the sub-THz radio flux}

\begin{figure}\centering
\includegraphics[width=0.7\linewidth]{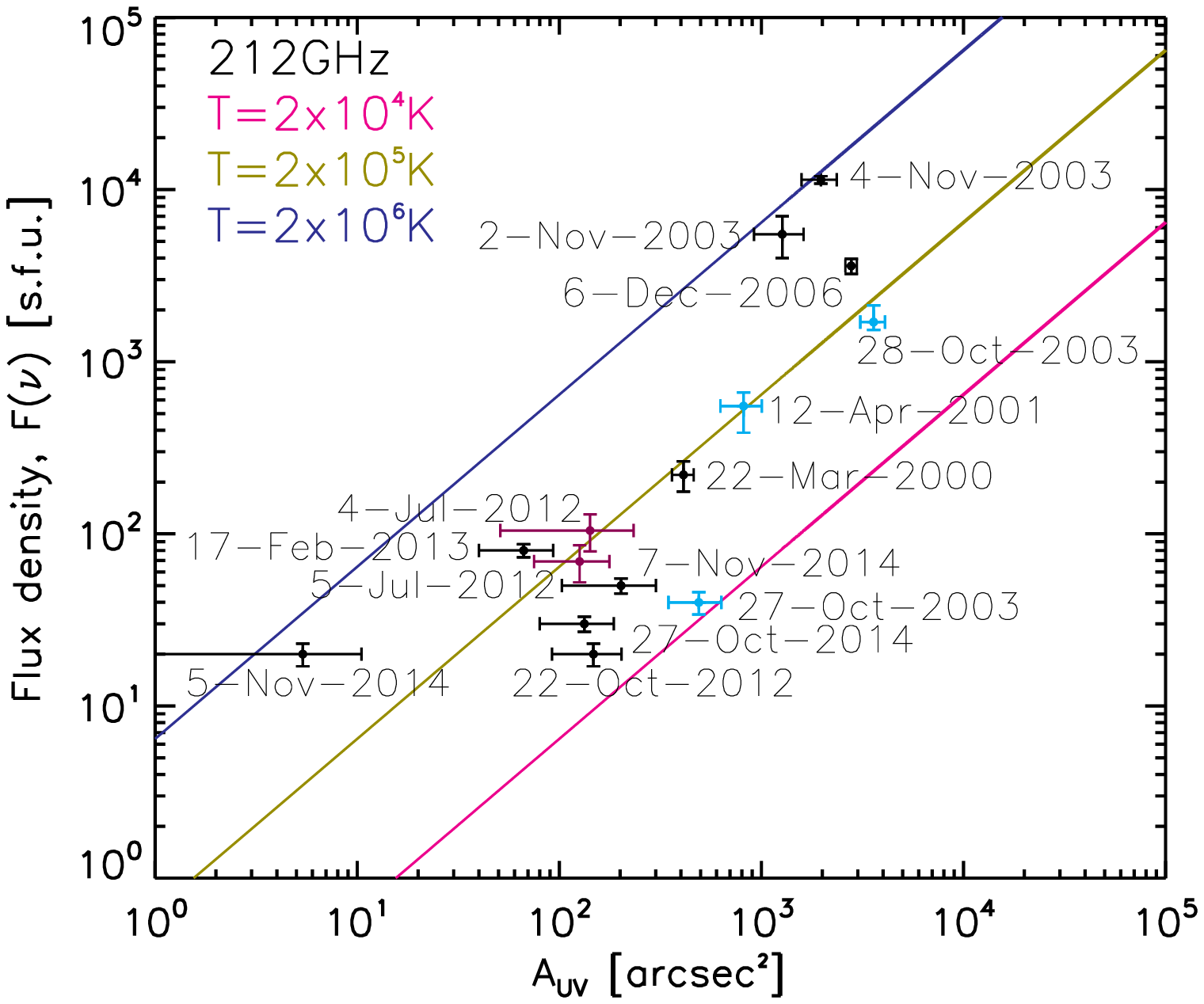}
\includegraphics[width=0.7\linewidth]{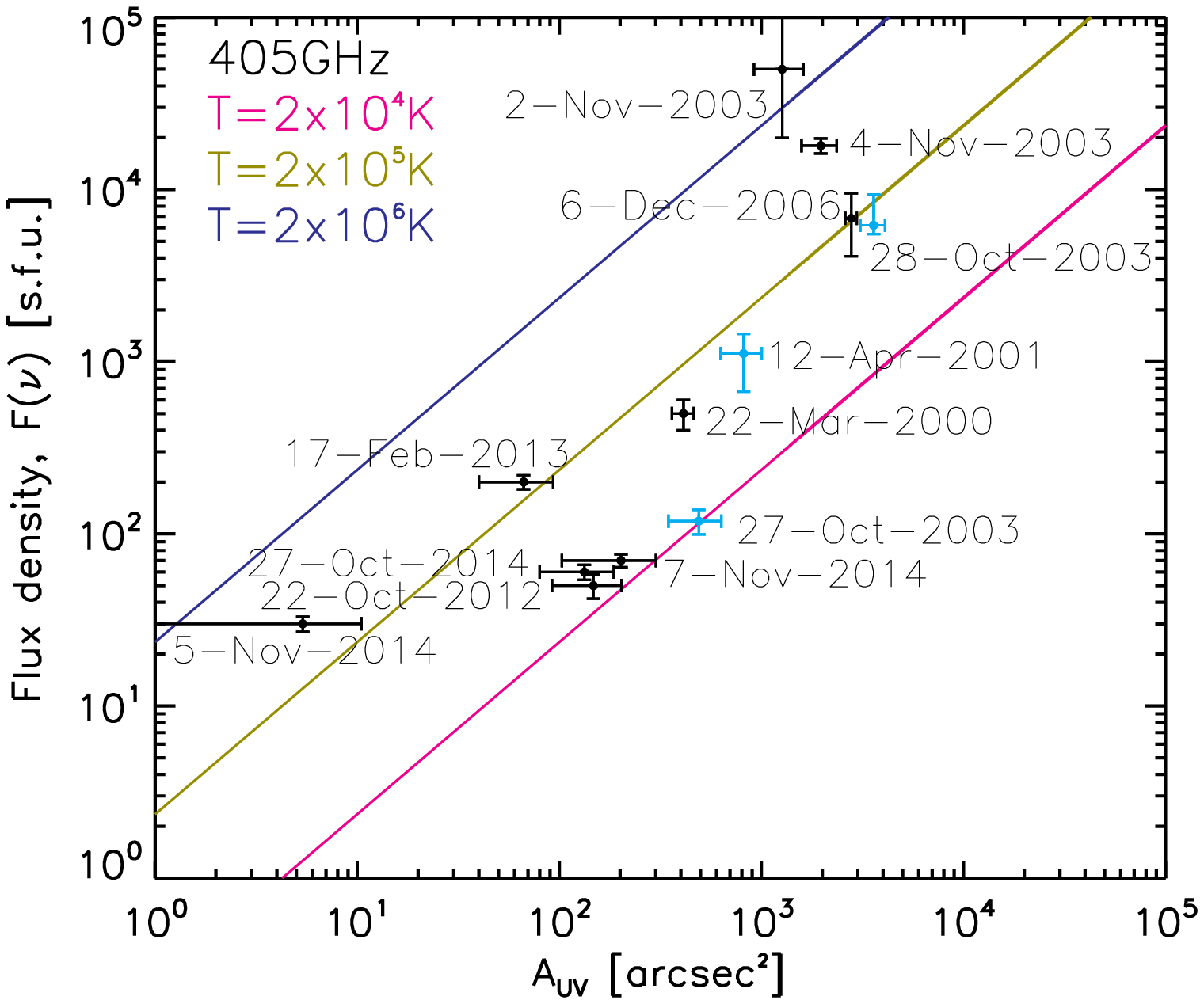}
\caption{\label{Fig_area_trace} Spectral flux density $F(\nu)$ versus UV ribbon area $A_{\rm{UV}}$ for 212~GHz (top panel) and 405~GHz (bottom panel) for flares with a positive spectral slope (9 events, black crosses and dates). For flares recorded at 230 and 345 GHz the fluxes were recalculated as  $F_{\text{230 GHz}} / (230^2)\times(212^2)$  and $F_{\text{345 GHz}} / (345^2)\times(405^2)$ respectively (3 events, blue crosses and dates). For two flares registered at 93 and 140 GHz the fluxes were recalculated in a similar way to $\nu=21$2 GHz and are shown only in the upper panel (2 events, purple crosses and dates).
 The flux density (solid lines) predicted by black-body emission for temperatures $T=2\times 10^4$, $2\times 10^5$, $2\times10^6$ ~K are shown by pink, dark yellow and dark blue lines respectively.}
\end{figure}

The observed spectral flux density $F({\nu})$ [1 s.f.u. = $10^{-22}$ W/m$^2$/Hz] is proportional to the area of the emitting source
and it is given by the Rayleigh-Jeans relation
\begin{equation}
F=\frac{2\nu^2k_BT}{c^2}\frac{A}{R^2}\simeq
722\left(\frac{\nu}{\mbox{100 GHz}}\right)^2
\left(\frac{T}{\mbox{$10^6$~K}}\right)
\left(\frac{A}{\mbox{$10^3$ arcsec$^2$}}\right) \mbox{ [s.f.u.]}\;,
\label{eq:flux_tb}
\end{equation}
where $k_B$  is the Boltzmann constant, $\nu$ is the frequency,
$c$ is the speed of light, $T$ is the temperature,
$A$ is the projected area of the radio source,
and $R$ is the Sun-Earth distance.
Therefore, area $A$ (Equation \ref{eq:flux_tb})
is an important parameter for a thermal emission model.
If the sub-THz emission originates from optically thick thermal
plasma in the upper chromosphere/transition region,
the area of the heated plasma should be proportional
to the radio flux $F\propto A_{\rm{UV}}$.
Figure \ref{Fig_area_trace} shows the dependence of radio flux
density $F({\nu})$ as a function of area $A_{\rm{UV}}$
for 212 and 405~GHz, 
for the 14 solar flares with a positive spectral slope (Table \ref{tab:list1}).
The area uncertainties for TRACE and AIA correspond to the maximum and minimum areas.  Figure \ref{Fig_area_trace} demonstrates that all radio-fluxes between 200-400 GHz can be explained by radiation from an optically thick
plasma with a temperature between $2\times 10^4$~K and $2\times 10^6$~K,
i.e. the plasma temperatures typical for the transition region.

\subsection{Thermal optically thick emission}

\begin{figure}\centering
\includegraphics[width=8cm]{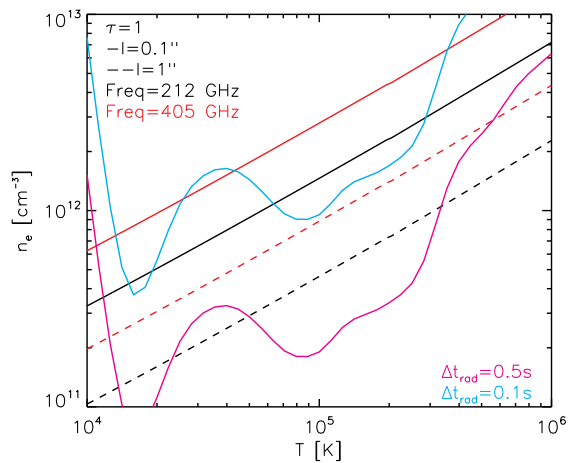}
\caption{\label{fig:rad_loss} Electron number density $n_e$ versus temperature for $T=10^4-10^6$ $K$, for $\tau=1$ and a geometrical depth of $l=0.1''$ (solid lines) and $l=1''$ (dashed lines) for $\nu=212$ (\textit{black})
and $405$~GHz (\textit{red}). The pink and blue lines indicate $n_e$ for the radiative loss times of $\Delta t_{rad}=0.5, 0.1$~s respectively.}
\end{figure}

The observations demonstrate that
flares with larger radio fluxes tend to have larger UV ribbon areas
(Figure \ref{Fig_area_trace}).
In order to explain the highest observed
radio flux density of $\sim 10^4$ s.f.u., the temperature of the plasma
should be $10^5-10^6$~K, even for the brightest flares
in October 2003 (Figure \ref{eq:flux_tb}).
The radio flux increasing with increasing frequency  would
require an optically thick source.
The optical depth $\tau_{\nu}$ over distance $l$ can be written as
\begin{equation}
\label{eq:tau_nu}
\tau_{\nu}=\frac{Kn_e^2}{T^{3/2}\nu^2}l\, ,
\end{equation}
where $K=9.78\times10^{-3}\times(18.2+\ln T^{3/2}-\ln\nu)$ for $T<2\times10^5$~K and $K=9.78\times10^{-3}\times(24.5+\ln T-\ln\nu)$ for $T>2\times10^5$~K \citep{1985ARA&A..23..169D}.
The above equation can be rewritten for an optical depth of $\tau_{\nu}=1$ to
express the density $n_e$ required to provide optically thick emission
\begin{equation}
n_e=\sqrt{\frac{T^{3/2}\nu^2}{Kl}}.
\end{equation}
For example, if the transition region is only 0.1~arcsec ($\sim$72 km) thick,
the plasma at a temperature of $10^5$~K which is optically thick at 405~GHz should have density of $3\times 10^{12}$~cm$^{-3}$ (Figure \ref{fig:rad_loss}).

\subsection{HXR flare footpoints and sub-THz emission}
RHESSI observations provide energy-dependent locations of HXR footpoints.
HXR sources are found  \citep[][]{2008A&A...489L..57K,2010ApJ...721.1933S,2010ApJ...717..250K}
at heights around $0.7-1.7$~Mm above the photospheric level. 30~keV HXR emission is observed to be produced at heights
of $\sim$1.7~Mm \citep{2008A&A...489L..57K},
where the plasma density exceeds $10^{12}-10^{13}$~cm$^{-3}$,
and co-spatial with white-light emission indicating ionisation caused by the
non-thermal electrons \citep{2011A&A...533L...2B,2012ApJ...760..142B}. These observations suggest that sub-THz emission is located in a region
where non-thermal electrons with energy $<30$~keV are depositing their energy
and the plasma densities are sufficient to make 0.4~THz emission optically thick.
The observation of ribbon-like HXR emission \citep{2007ApJ...658L.127L}
is often viewed as an additional indication that UV emission
is caused by the energy deposition of HXR producing electrons.

\subsection{Time variability of sub-THz emission}
Relatively dense plasma $n > 10^{11}$~cm$^{-3}$ heated by energetic electrons to temperatures 0.1-1~MK leads to enhanced radiation \citep{1976ApJ...204..290R,2011ApJ...735...70J,2016ApJ...833..184S},
so that the radiation losses would lead to effective cooling.
The radiative cooling time is \citep{1969ApJ...157.1157C,1976ApJ...204..290R,2016ApJ...833...76B}
\begin{equation}
\Delta t_{rad}=3 \frac{k_BT}{n_e\Lambda},
\end{equation}
where the radiative loss function is calculated using the CHIANTI \citep[][]{1997A&AS..125..149D} function rad\_loss.pro\footnote{\url{http://www.chiantidatabase.org/tech_reports/09_rad_loss/chianti_report_09.pdf}}. The radiative loss can be estimated using the approximation
$\Lambda (T)\simeq 1.2\times10^{-19}T^{-1/2}$, where $T$ has units of Kelvin and $\Lambda (T)$ has units of [erg~cm~$^{3}$~s$^{-1}$],
giving $\Lambda (T)\simeq 4\times 10^{-22}$ for $T=10^5$~K.

The radiative loss time for plasma temperatures between $T=10^4-10^6$~K
is shown in Figure~\ref{fig:rad_loss}. The results suggest that the plasma
can quickly cool if the heating time is greater than the radiation loss time.
The observations by \citet{2009ApJ...697..420K} show that the flux changes
with a characteristic time of $0.1-1$~s. This suggests
that the interplay between non-thermal electron heating and the
radiative cooling of dense plasma can explain
the observed variability of sub-THz emission.

Quasi-synchronized broadband solar flare rapid pulsations at sub-second time scales are observed at HXR and cm-mm wavelengths \citep{1983Natur.302..317T,1984ApJ...279..875C,2000SoPh..197..361K}.
These short-time scale features are often called `elementary bursts' following \citet{1978SoPh...58..127D}. In particular, efforts to search for small scale structures have also been made in optical wavelengths, typically
$H_\alpha$ observations \citep{2000ApJ...542.1080W,2000A&A...356.1067T,2000ASPC..206..426K,2008ApJ...677.1367A,2016ApJ...822...71F}
For example, \citet{2000ApJ...542.1080W}
reported fine temporal structures of the order $0.3-0.7$~seconds
in $H_\alpha$ off-band emission, and it is found that these fine structures
only occur at flare kernels whose light curves are well correlated
with HXR emission. This close correlation
suggests that the sub-THz pulsations are caused
by fast electron energy deposition or by thermal conduction
in the transition region and heated chromosphere.
In such a scenario,  the finite temperature of the ambient plasma
plays an important role in determining the energy deposition \citep{2015ApJ...809...35K,2015JPhCS.642a2013J}.
An alternative way proposes heating by fast electrons accelerated in situ in the chromosphere due to the magnetic Rayleigh-Taylor instability \citep{2015SoPh..290.3559Z,2017SoPh..292..141Z}.

\section{The sub-THz emission model}

During solar flares, the transition region defined as the plasma between the chromosphere and corona \citep{1983SoPh...86..159T},
has a complicated structure.
It can be viewed as an envelope covering the cool evaporating chromospheric plasma \citep{1983SoPh...86..159T}.
This can be seen in limb flares, where the dense chromospheric plasma
expands into corona, so the `transition region' becomes vertically
extended to coat this plasma. Moreover, a substantial vertical
enhancement of the plasma density is required
to explain RHESSI observations \citep{2010ApJ...717..250K}
and is visible in chromospheric evaporation simulations \citep[e.g.][]{1980ApJ...242..336M,2009CEAB...33..309K}.

While the spectrum decreasing with frequency is a continuation of
the standard gyrosynchrotron spectrum into the sub-THz range \citep{1985ARA&A..23..169D}, the frequency-growing spectrum
requires a different mechanism. For example,  SOL2001-08-25T16:45
observed by \citet{2004SoPh..223..181R} gives 
a large radio flux of 5500~s.f.u. at 400~GHz, 
but it is consistent with the gyrosynchrotron spectrum decreasing
with frequency that hides thermal emission.
As a model to explain the radio emission growing with frequency,
we propose that the large fluxes of the sub-THz emission
are due to the large areas of these flare ribbons.
Therefore, the sub-THz flare component is produced at the heated flare ribbons
and flares with small ribbon areas should produce weaker
sub-THz emission, while the flares demonstrating extended flare ribbons should
be strong sub-THz emitters. Then, the thermal emission from an optically thick
transition region and low coronal plasma, with temperatures between 0.1-2~MK
will produce a spectrum growing with frequency $F(\nu)\propto \nu^\alpha$,
where $\alpha \leq 2$.

In this scenario, the energy deposition rate is balanced
by effective thermal radiative losses. As the plasma cools down quickly,
the radio emission will stop as soon as the energy supply has ended.
Variations in the electron acceleration rates will also
produce radio flux variations at the time scales of radiation
cooling ($0.1-1$~s) and provide a viable explanation for the sub-second
variations of the flux \citep{2009ApJ...697..420K}.

Our model predicts that when the flare is compact
(i.e. its UV ribbons have a small area) then the sub-THz radio flux is also small.
We found observational support of this
prediction. Table \ref{tab:list2} shows that powerful GOES X-class flares
can be weak emitters of sub-THz radiation. Although the flares
have a GOES class of M and X, they are weak sub-THz flares.
The peak of their radio flux is of the order of $\sim 100$~s.f.u.,
and they have small ribbon areas not exceeding $500$~arcseconds$^2$.
Figure~\ref{Fig_area_trace} also demonstrates that 
there is a smooth transition from flares with weak sub-THz emission to
flares where the 400~GHz flux density exceeds $10^4$~s.f.u.,
suggesting the same emission mechanism for all flares.

It is interesting to note that the model is also supported
by observations of transition region lines.
UV observations of the transition region during flares show good correlation
with HXR emission, so that individual peaks in HXRs,
can be identified with individual peaks of UV emission \citep{1983SoPh...86..159T,2001ApJ...560L..87W}.
The ratios of transition region lines indicate \citep{2007ApJ...659..750R}
that the emitting region is about 100~km thick for densities of the order
of $10^{12}$~cm$^{-3}$, similar to the densities required for the
sub-THz emission, see Figure~\ref{fig:rad_loss}.

\section{Summary}

We have examined solar flares that produce strong sub-THz emission
rising with frequency. Flares with strong radiation in the sub-THz frequency range
tend to have large and extended flare ribbon areas, clearly seen in UV.

To explain the observations, we propose that the rising component of the sub-THz radiation is produced via thermal bremsstrahlung (free-free emission) from an optically thick plasma located in the transition region heated by precipitating non-thermal electrons, ions or heat conduction. Due to flare heating
of the chromosphere, the plasma evaporates and fills the loop,
forcing the transition region to lower heights \citep{1974SoPh...34..323H}
and hence higher density regions  \citep{1980ApJ...242..336M,2009CEAB...33..309K},
where radio emission with  $\nu <0.4$~THz emission becomes optically thick.
Plasma in the range of temperatures of 0.1-1~MK then
produces enhanced sub-THz emission.
The model provides a frequency spectrum that grows with frequency,
for flares with ribbon areas
of $\sim 1000$~arcsec$^2$, that produce a radio flux density of $10^4$~s.f.u. \citep{2012ASSP...30...61K}.
We note that the quiet-Sun transition region is located
at an altitude of $\sim 2$~Mm with a plasma density
of $10^{11}$~cm$^{-3}$ \citep{1981ApJS...45..635V} and it
is optically thin to such radio emission, so the sub-THz emission is produced
in the chromosphere where the plasma temperature is $<10^4$~K.
It should be noted that in the presence of strong gyrosynchrotron emission,
the flux in the sub-THz range could be dominated
by this gyrosynchrotron component producing smaller radio source sizes.
In particular, the observations by \citet{2004A&A...420..361L} are consistent
with the gyrosynchrotron source dominating 210~GHz observations,
while the gyrosynchrotron emission dominates up to 400~GHz in SOL2001-08-25.

The rapid heating by non-thermal electrons with energy $\sim 20$~keV
and radiative cooling of the dense $>10^{12}$~cm$^{-3}$
plasma is likely to produce variations at sub-second time scales
and this should be temporary correlated with the HXR flux of electrons
and/or ions, the particles that produce evaporation
and hence an enhanced density of heated evaporating plasma.

In summary, we have analysed virtually all flares
with a rising sub-THz component reported in the literature.
We analysed UV ribbon areas, a signature of energy deposition in the chromosphere, and the observational results suggest that the strong sub-THz emission is associated with large UV ribbon areas.
Large flare ribbon areas can explain this sub-THz emission
as the thermal emission from optically thick plasma with a temperature of 0.1-2~MK. Our model is consistent with the standard solar flare scenario
and shows that the sub-THz emission
could be a valuable temperature diagnostic tool of the dynamic transition
region during solar flares.

\begin{acknowledgements}
      EPK is indebt to P. Kaufmann for inspiring discussions and hospitality
      in S\~{a}o Paulo, Brazil. EPK \& NLSJ gratefully acknowledge the financial support from the STFC Consolidated Grant ST/L000741/1. GM is supported by the Russian Science Foundation (project no. 16-12-10448).
\end{acknowledgements}

 \bibliographystyle{aa} % style aa.bst
\bibliography{refs} % your references Yourfile.bib

\end{document}